\documentclass[conference]{IEEEtran}
\IEEEoverridecommandlockouts
\usepackage{cite}
\usepackage{amsmath,amssymb,amsfonts}
\usepackage{algorithmic}
\usepackage{graphicx}
\usepackage{textcomp}
\usepackage{xcolor}
\usepackage{subfig}
\usepackage{url}

\usepackage{fancyhdr}
\begin{document}

\title{
Architectural Form and Affect: \\A Spatiotemporal Study of Arousal

\thanks{
The research leading to these results has received funding from the European Union H2020 Horizon Programme (2014-2020) under grant agreement 952002, project PrismArch: Virtual reality aided design blending cross-disciplinary aspects of architecture in a multi-simulation environment.
}
}

\author{\IEEEauthorblockN{Emmanouil Xylakis}
\IEEEauthorblockA{\textit{Institute of Digital Games} \\
\textit{University of Malta}\\
Msida, Malta \\
emmanouil.xylakis@um.edu.mt}
\and
\IEEEauthorblockN{Antonios Liapis}
\IEEEauthorblockA{\textit{Institute of Digital Games} \\
\textit{University of Malta}\\
Msida, Malta \\
antonios.liapis@um.edu.mt}
\and
\IEEEauthorblockN{Georgios N. Yannakakis}
\IEEEauthorblockA{\textit{Institute of Digital Games} \\
\textit{University of Malta}\\
Msida, Malta \\
georgios.yannakakis@um.edu.mt}
}

\maketitle
\thispagestyle{fancy}

\begin{abstract}
How does the form of our surroundings impact the ways we feel? This paper extends the body of research on the effects that space and light have on emotion by focusing on critical features of architectural form and illumination colors and their spatiotemporal impact on arousal. For that purpose, we solicited a corpus of spatial transitions in video form, lasting over 60 minutes, annotated by three participants in terms of arousal in a time-continuous and unbounded fashion. We process the annotation traces of that corpus in a relative fashion, focusing on the direction of arousal changes (increasing or decreasing) as affected by changes between consecutive rooms. Results show that properties of the form such as curved or complex spaces align highly with increased arousal. The analysis presented in this paper sheds some initial light in the relationship between arousal and core spatiotemporal features of form that is of particular importance for the affect-driven design of architectural spaces.
\end{abstract}


\section{Introduction}\label{sec:introduction}

Architects are required to consider three core Vitruvian criteria during the design process: the \textit{utility} (i.e. function), the \textit{solidity} (i.e. material) and the \textit{beauty} (i.e. aesthetics) of the built outcome \cite{rowland2001vitruvius}. The contemporary approach to architecture largely follows two views on this topic: one defining function as the main form-giver \cite{Sullivan1896}, the other placing the human at the center of design \cite{ruta2019comparison}. This latter view gave birth to approaches aimed at understanding the human psyche and delivering affective experiences that promote well-being and quality of living \cite{smith2012interior, kirillova2020workplace, petermans2014design, de2019tutorial}. However, quantifying the impact of architectural design on the human subject and its potential to deliver satisfying experiences is a great challenge, as the affective response to architectural design is subject to personal taste, values, cultural influences, aesthetics, and expectations.

To understand how design elements impact our affective response, a common practice in architectural design research is the identification, definition and adaptation of the parameters that comprise the built environment. By doing so, the relations between the features involved and the emotions they elicit could arise and be quantified. The capability of \emph{form} to elicit emotions, however, still lacks sufficient evidence compared for instance to lighting and color \cite{banaei2020emotional}.

Motivated by the lack of a comprehensive study on the impact of architectural form on emotions, this paper leverages methods and theories taken from environmental psychology, and best practices for affect annotation and modeling \cite{picard2000affective} to shed light on the \emph{spatiotemporal} relationship between affect and form. In particular, the objective of this study is three-fold. First, to identify the most prevalent design features and spatial properties in the literature in terms of their impact on affect. Second, to create an affective corpus of spatial transitions and corresponding time-continuous annotations of arousal. Third, to analyze the impact of different design features on affect by treating both the design features and the annotations in a relative manner \cite{yannakakis2017ordinal,camilleri2017generalmodels}. 
This paper contributes to an extensive body of work on the impact of space, form, and color on affect, and is the first instance of time-continuous affect annotation on a \emph{spatiotemporal} navigation task. The paper also proposes and tests two ways of treating arousal in a relative fashion, based on comparisons between mean arousal values in consecutive spaces and based on the degree of change of arousal during the \textit{arrival} into a new space. Both treatments are able to capture strong effects between a room's curvature and perceptual complexity on arousal.

\section{Related Work}\label{sec:related}

Studying the affective response of humans to the built environment is undeniably challenging, as it is arguably difficult to compartmentalize which elements are present or which elements draw our attention (and to what extent). Identifying the momentary emotional response that might be at play during the process of spatial evaluation is a field of study combining disciplines and theories coming from psychology, architecture, visual perception and cognition. Interestingly, the emotional effect of spaces has been well-studied in the domain of digital games \cite{niedenthal2009patterns,joosten2012influencing}. This is likely because it is easy to build many variations of the same space inexpensively, but also because games can elicit strong emotions through their gameplay, in-game events (e.g. enemies or sound effects), but also spaces. 

The \emph{curvature} of spatial elements has been investigated extensively in several studies, not only in terms of affective response but also as a parameter of objects' shape \cite{bertamini2021study,dazkir2009emotional,gomez2018preference}. As a property of any shape, designers' preoccupation with curvature is linked in many ways to our potential dislike of angularity and a sense of perceptual threat that it conveys \cite{ruta2019comparison}. Curvature as a parameter of form has been investigated as a decorative feature in interior openings or in fa\c{c}ade design \cite{ruta2019comparison,chamilothori2019effects}. In \cite{ruta2019comparison}, authors sought to study the potential link of artistic expertise to preference for curvature; 24 female participants were introduced to three different tasks---pairwise preference, multiple psychological variables ranking and preference ranking---in comparing four different types of fa\c{c}ades (curved, mixed, angular and rectilinear) projected on a wall. Results from that experiment confirmed a preference towards curved and mixed properties of fa\c{c}ade design. 
Rather than using self-reports, Shemesh \emph{et al.} \cite{shemesh2017affective} coupled electroencephalogram (EEG) data of 42 participants with virtual spatial stimuli from within a VR headset. Results showed that there were consistent differences in terms of EEG responses between experienced and inexperienced designers to properties of curvature, irregularity and rectilinearity. A supplementary test based on rankings showed a preference for curved spaces by participants inexperienced to design; experienced participants however tended to prefer rectilinear spaces. 
Banaei \emph{et al.} \cite{banaei2017walking} compared two methods for measuring arousal in interior scenes, via the self assessment manikin (SAM) \cite{bradley1994measuring} and via EEG. SAM tests confirmed positive correlations between arousal and curvature and negative correlation with rectilinear and angular spatial elements.
In another study using biofeedback, functional magnetic resonance imaging (FMRI) was used to study the effect of interior scenes with spatial properties related to curvature and rectilinearity, varying ceiling height and either open or enclosed arrangement \cite{vartanian2013impact}. In that study, 18 participants indicated their preferred scenes and on a second test chose which scene to approach or avoid during an FMRI scanning procedure. The study indicated a tendency towards curved over linear arrangements.

\emph{Ceiling height} and \emph{volume} of an interior space can convey feelings of freedom or confinement, and have been primarily investigated in studies regarding memory and attention. Meyers-Levy and Zhu studied the effects of ceiling heights on participants in delivering a series of tasks \cite{meyers2007influence} within physical  environments. In a first experiment, 32 participants responded to six Likert-scale questions---reflecting on freedom-related feelings and confinement-related body states---and performed tasks of solving 12 anagrams. In a second experiment, 100 participants performed a categorization task and a product evaluation task. Results validated the authors' hypothesis that higher ceilings result in a higher feeling of freedom; moreover, low ceilings prompt relational (item-specific) processing, while high ceilings prompt abstract ideation. The capacity for larger volumes to trigger a higher perceived arousal is explained by Niedenthal \cite{niedenthal2009patterns} where the experience of awe is conveyed by the Gothic setting of churches and castles in the game \textit{Resident Evil 4} (Capcom, 2005). The change of volume either from low-ceiling rooms to higher or the opposite is a tool that in many cases is employed by video game designers and architects in order to accentuate a change in spatial relationship, usually from a transitional ``no-space'' to a space; this process is identified as an \textit{arrival} \cite{totten2014architectural}.

\emph{Lighting} and its effect on human experiences has been frequently tested within virtual environments \cite{marples2020effect,chamilothori2019effects} or through rendered static images \cite{rockcastle2014measuring}. Illumination parameters such as brightness, color and luminance distribution are studied frequently \cite{joosten2012influencing} in terms of affective responses, preference and attention tasks \cite{marples2020effect,rockcastle2017experiment}. The effects of illumination is also well-studied in video game research \cite{graja2020impact,el2009dynamic}, primarily regarding player performance and gameplay. In multiple studies \cite{knez2008lighting,marples2020effect} the impact of ambient light color was studied in terms of completion time for a maze navigation task. In both studies, blue ambient color resulted in longer completion times compared to red ambient color, although the completion times under neutral light (compared to the other two treatments) differed between the two studies. Regarding the impact of color on affect, Joosten \emph{et al.} \cite{joosten2012influencing} investigated the effect of ambient lighting color on the player's self-reported pleasure, arousal and dominance \cite{mehrabian1996pleasure} levels during the completion of a task for a custom-made module in the game \textit{Neverwinter Nights} (Bioware, 2002). Responses from 60 participants, each conducting three playthroughs of four rooms per playthrough, showed that red ambient color scored highest in terms of arousal and yellow scored highest in terms of valence among experienced players. This study highlighted the importance of color and perceived affect, and how it can be influenced by the players' level of experience.

Several studies explore more than one design feature in tandem. Ergan \textit{et al.} \cite{ergan2018towards} examined how certain spatial features could stand out from their surroundings and affect the human experience. Through dual image comparison of generated scenes, features were put in polar opposites and compared. Features explored in this study include daylight, openness of a space, ceiling height, level of artificial lighting, symmetry of interior elements and contour curvature. Based on participants' preferences, most impactful features were the access to windows, space openness, ease of access, flexibility in isolation and color of surfaces. It is worth noting that two of the top features relate to how open or cramped a space is perceived. Banaei \textit{et al.} \cite{banaei2017application} sought to cluster different interior elements according their form and appearance. The authors collected a series of living room arrangements that followed different trends and defined 25 clusters that included features such as curvature, scale, location and angle. Results demonstrated a stronger presence of rectilinearity over curvature in the selected sample but also demonstrated the inability of some elements to be used for a categorization task.

The aforementioned studies highlight the current state of affective computing in relation to form and spatial affect. The reviewed literature shows a variety of perspectives, hypotheses, and ways of evaluating affect, but also considers a broad range of design parameters. This rich body of work, however, appears to lack the temporal dimension both in the treatment of data but also regarding how spatial stimuli are presented for annotation. The importance of the reactive nature of architecture and space appraisal is highlighted within this study and the method for eliciting and capturing such time-sensitive data is discussed below.

\begin{figure}
\centering
\subfloat[]{\includegraphics[width=0.45\linewidth]{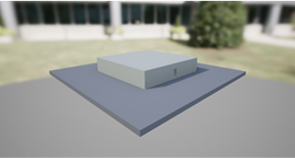}\label{fig:model_00}}\quad
\subfloat[]{\includegraphics[width=0.45\linewidth]{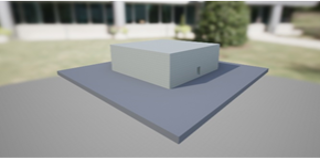}\label{fig:model_01}}\\
\subfloat[]{\includegraphics[width=0.45\linewidth]{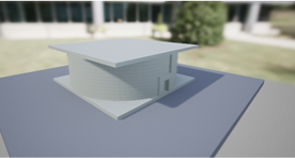}\label{fig:model_10}}\quad
\subfloat[]{\includegraphics[width=0.45\linewidth]{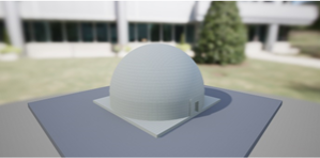}\label{fig:model_11}}
\caption{External view of rooms for combinations of size and curvature: large size (right column), curvature (bottom row)}
\label{fig:4_models}
\end{figure}

\begin{figure*}
\centering
\includegraphics[width=0.95\linewidth]{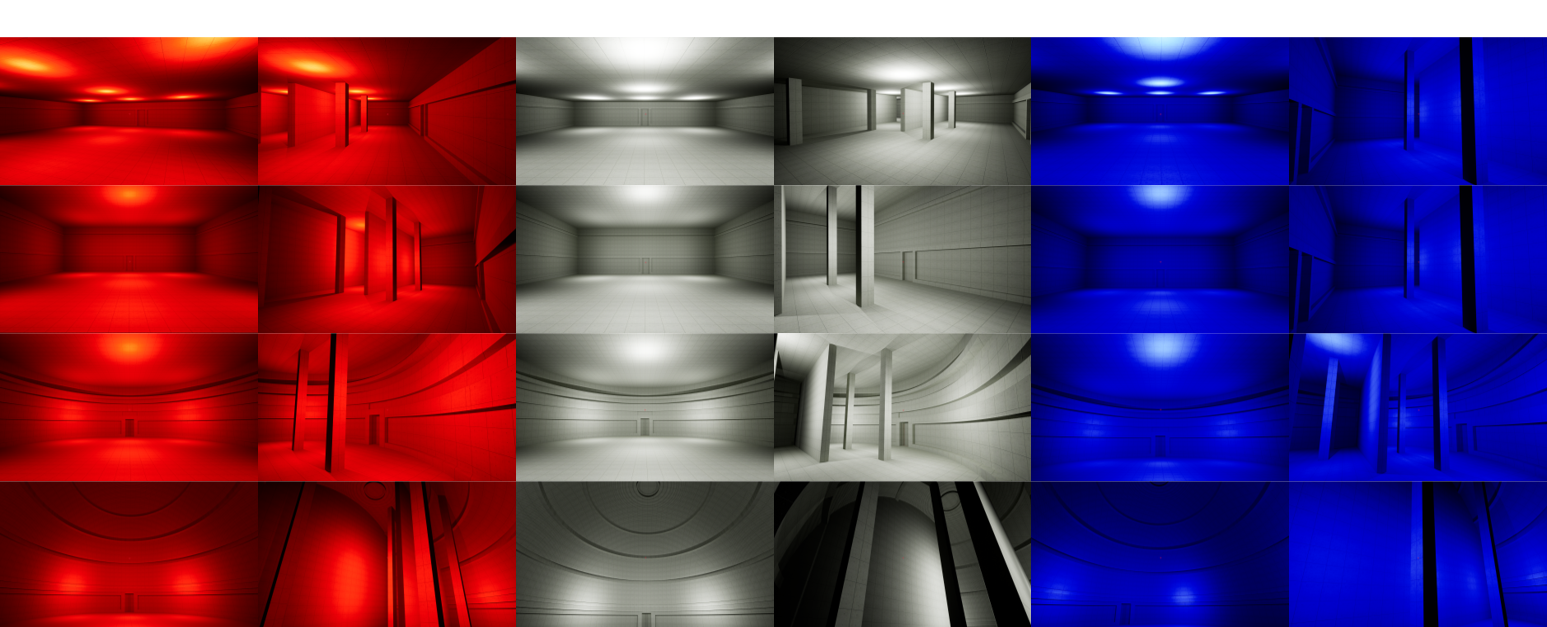}
\caption{Views of the 24 rooms examined in the AffRooms corpus}
\label{fig:all_rooms}
\end{figure*}

\section{Experimental Setup}\label{sec:methodology}

To explore the affective responses that different properties of architectural form and illumination may elicit, we conducted a user study with a broad range of spaces and spatial transitions. This section describes the elicitors (i.e. the video dataset of spatial navigation) and the annotation methods used for building our affective corpus. The video dataset and annotations (raw and processed) are available in a public repository\footnote{\url{https://drive.google.com/drive/folders/1R4XhMYKcULJprt_OCdS0ysptOuEDqURj?usp=sharing}}.

\subsection{Features Explored}\label{sec:features}

Four design features were selected based on their contribution to the appearance of the space, either in relation to the \textit{room geometry} or \textit{ambient lighting}.

In terms of room geometry (form), we explore the effect of room contour \textbf{curvature}, interior \textbf{complexity}, and the room's \textbf{size}. Each of these properties were either present or absent in each room. In the case of room size, low ceilings represent small size (absence) and high ceilings represent large size (presence); Figure \ref{fig:4_models} shows all possible room sizes. Contour curvature is explored as a potential contributor to arousal, compared to rooms that are defined by rectilinear contours. Different combinations of curvature and size affect the room layout differently: small size and high curvature result in cylindrical rooms (Fig. \ref{fig:model_10}) while large size and high curvature result in a dome-like structure (Fig. \ref{fig:model_11}). Interior complexity introduces boundaries and obstructions during navigation: complex spaces have symmetrically placed columns and two walls in the middle of the room that obstruct both visibility and the path of the user navigating through the room. 

Following the paradigm of previous studies on the effect of color in navigable spaces \cite{niedenthal2009patterns}, we explore illumination with red, blue and white as the possible colors of each room. Light sources are distributed within each room so as to provide general illumination from multiple sources and remain the same for all rooms. In our analysis, we use color warmth (1 for red, 0 for white, -1 for blue) to track changes in illumination.

With three features of room geometry with two possible states each, and a color illumination feature with three states, the possible combinations of rooms are 24. Each room has the same exterior dimensions, consisting of 20 meters width by 20 meters depth. Depending on the size and curvature parameter combinations, heights range from 3 to 12 meters. The final rooms, from a first-person view, are shown in Figure \ref{fig:all_rooms}.

\subsection{The AffRooms Corpus}\label{sec:dataset}

\begin{figure}
\centering
\includegraphics[width=0.95\linewidth]{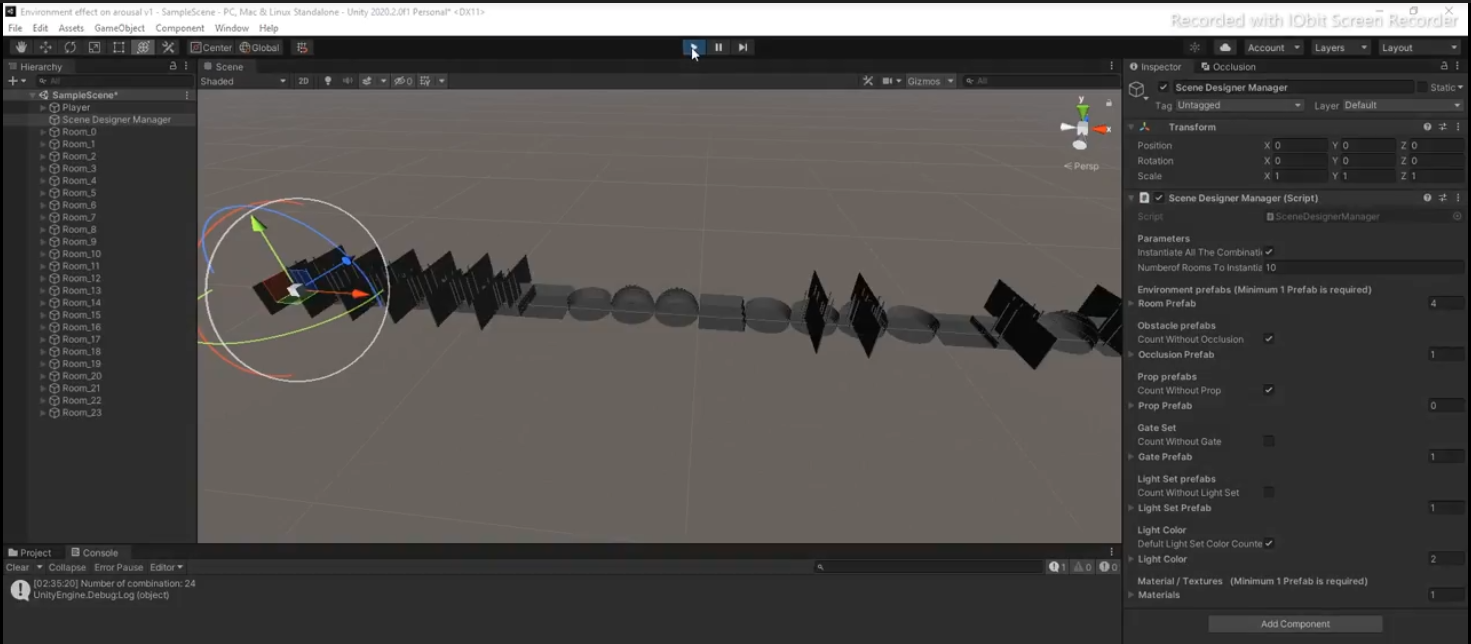}
\caption{Sequence of rooms generated in the Unreal Engine}\label{fig:ue4_recording}
\end{figure}

The goal of the experiment is to capture affect annotations during navigation and examine how each design feature impacts our perceptual arousal in a temporal manner. To increase the consistency of the annotators' experience, all annotators received the same elicitors in the form of pre-recorded footage of the first author navigating through a sequence of rooms. To provide a diverse set of spatial transitions from one room to the next, multiple sequences of rooms were produced and navigated through to provide multiple videos.

The sequence of rooms was generated randomly within Unreal Engine (see Fig.~\ref{fig:ue4_recording}). In each generated sequence, all 24 possible rooms appear once. All 24 rooms are placed in a navigable sequential manner along a straight path, each separated with a sliding door. The player starts in a small entrance room, before they pass through the first sliding door to view the first room. For each recorded playthrough we ensured that while navigating and capturing, the camera's field of view would capture each room's distinct features. During the navigation tasks, the player's position and angle of the camera viewport were periodically logged, as was the time stamp when the player entered a new room and the type of features of each room. This allows us to align the arousal annotations and match them to the design features, as discussed in Section \ref{sec:processing}.

Twenty random sequences of rooms were generated and navigated, resulting in 20 playthrough videos that capture a diverse set of spatial transitions. The average duration of each video is 186 seconds (ranging from 164 to 240 seconds). 

\subsection{Annotation of Arousal}\label{sec:annotation}

\begin{figure}
\centering
\includegraphics[width=0.95\linewidth]{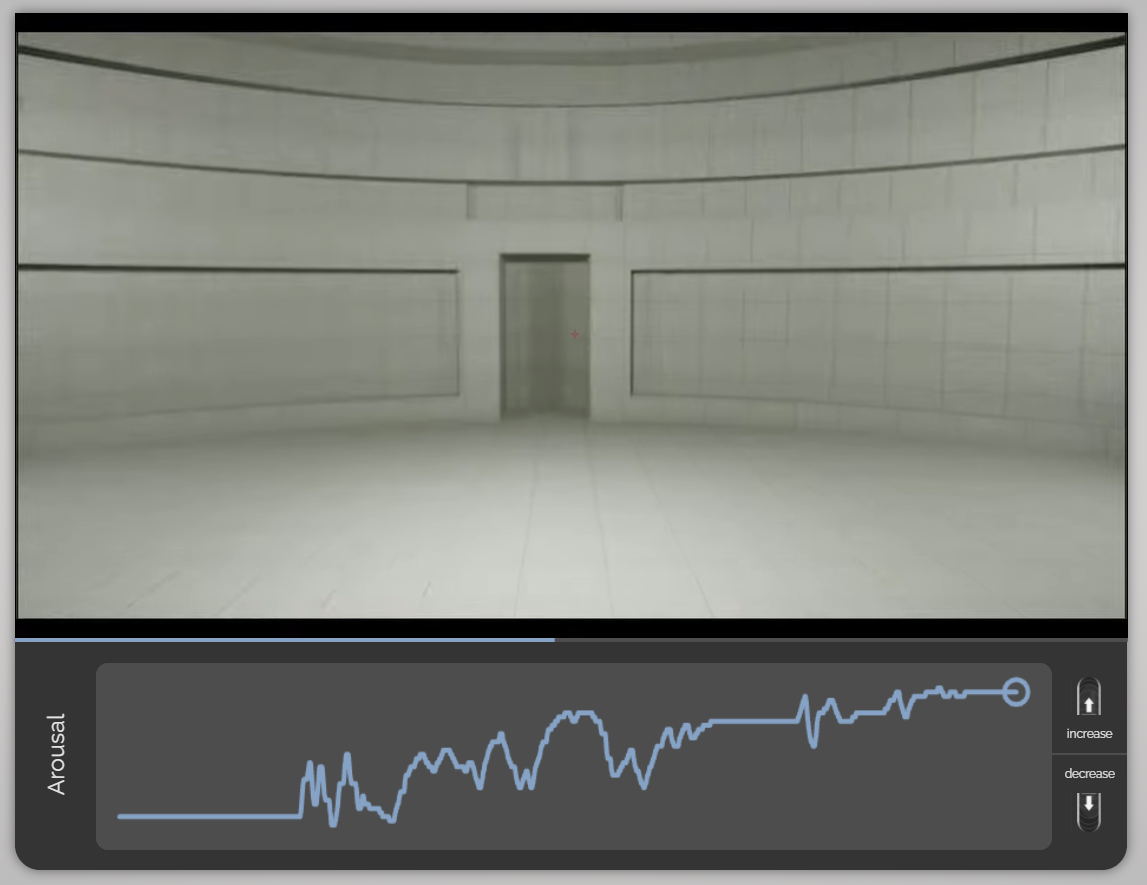}
\caption{Screenshot of PAGAN during an annotation task: navigation video (top) and continuous arousal annotation (bottom)}\label{fig:PAGAN_annotation}
\end{figure}

This study uses a continuous and unbounded method to capture annotators' reactions while viewing a pre-recorded video of a spatial navigation task. Arousal data was collected through the PAGAN video annotation tool \cite{melhart2019pagan}, which allows users to report changes in a single affect dimension while they watch a video.  Annotation traces collected by PAGAN are aligned to the video frames. In this experiment, we use the RankTrace \cite{lopes2017ranktrace} annotation protocol, which allows users to define the degree of change of the affective dimension in an unbounded fashion. Figure~\ref{fig:PAGAN_annotation} shows an instance of PAGAN during annotation. Users can control the degree of change through their mouse wheel.
 
Three participants with experience in PAGAN and RankTrace annotation protocols were recruited to annotate the 20 videos of the AffRooms dataset. All participants are research staff of University of Malta with expertise on artificial intelligence, affective computing and digital games. All participants are male, aged between 22 and 36. All appropriate consent was provided by participants; data was collected online, stored in the PAGAN database, and no personal data was retained. Annotation was done by participants remotely, with no interaction with the authors and outside of a controlled lab environment. The study follows the view of arousal as affect intensity rather than physiological activation. Each participant was given the following definition of arousal:
\begin{quote}
\textit{Arousal in the context of spatial appraisal is defined here as the momentary amplitude of emotions elicited during this process. An environment characterized as having positive amplitude is an environment described as exciting, tense, stimulating, wakened and/or intriguing. An environment assessed with a negative amplitude in arousal is an environment evoking feelings of boredom, fatigue, flatness, tiresomeness, calmness and/or relaxation.}
\end{quote}
After the participant had read the arousal definition and some guidelines for interacting with RankTrace, annotations were performed in a single session. Each participant was presented a random ordering of the 20 videos. The whole session for each user took around an hour to complete. Users had the option to pause the session at any moment throughout.

\subsection{Processing Arousal}\label{sec:processing}

\begin{figure}
\centering
\includegraphics[width=\linewidth,clip,trim=0 0 0.4cm 0]{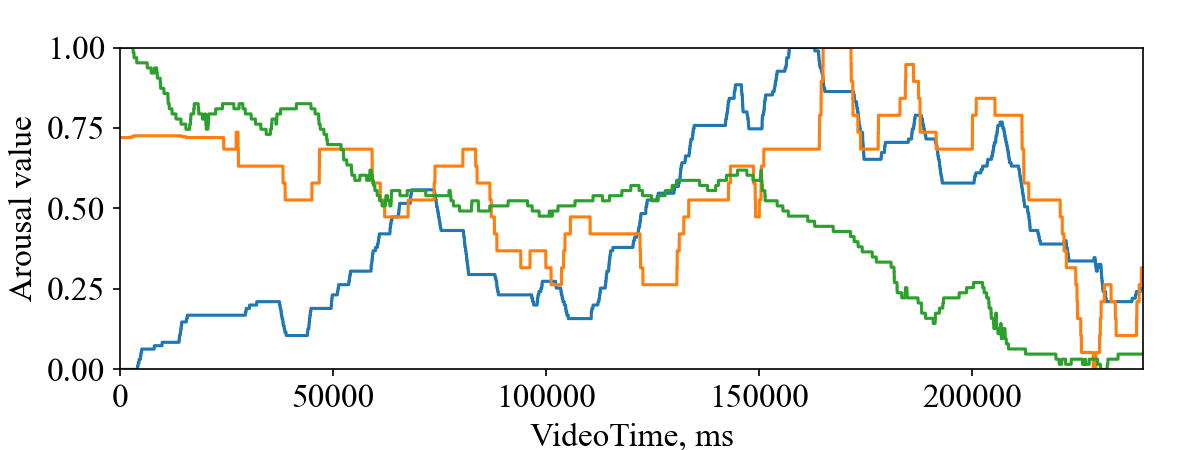}
\caption{Arousal traces of all three annotators, in different color, for the same video}
\label{fig:3_arousal_traces}
\end{figure}

Since each annotation of a video is unbounded, as a first step the arousal traces were normalized to $[0,1]$ via min-max normalization independently (i.e. on a per-user, per-video basis). Since PAGAN does not have a specified sampling rate, the trace is resampled at 1000Hz. Figure \ref{fig:3_arousal_traces} shows the normalized arousal annotations of the three participants on the same video. The annotation traces were processed in a relative fashion, in two different ways described below.


\subsubsection{Mean Room Arousal}
\label{sec:processing_mean}

\begin{figure}[!tb]
    \centering
    \includegraphics[width=\linewidth]{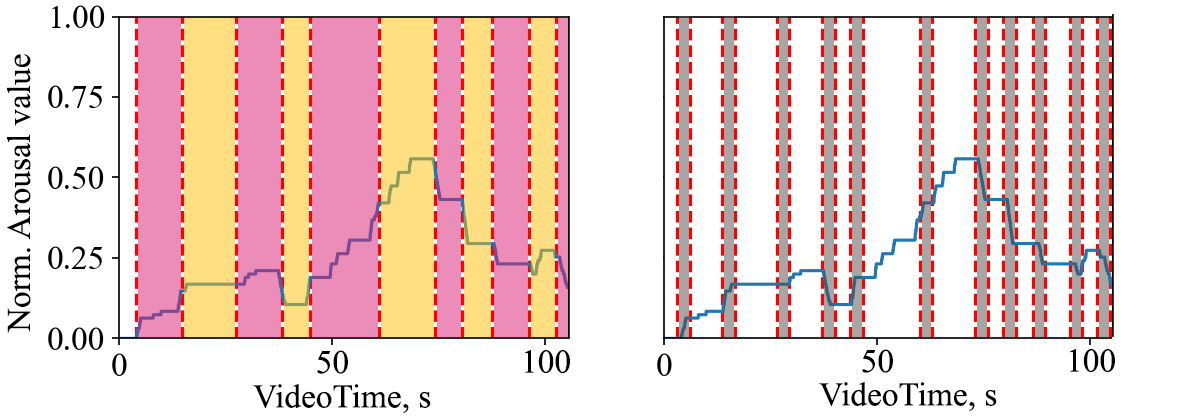}
    \caption{Segment of a user's annotation trace after processing. Left: trace split into room time windows (yellow and magenta). Right: arrival time windows (gray) extracted from trace.}
    \label{fig:room_window}
\end{figure}

A straightforward way to assess the impact of each spatial feature is to split the video playthrough based on when the player enters a new room, and average the annotated arousal per room. We identify the \textit{room time window} as the interval from the moment the room is entered until the moment the next room in the sequence is entered. Figure \ref{fig:room_window} shows the different room windows for one annotator's trace. We average the arousal values within that window to calculate the \textit{room's mean arousal} value. Comparing how changes in terms of each design feature match changes between rooms' mean arousal is a straightforward way to process the data as discussed in Section \ref{sec:results_changes}.

\subsubsection{Arrival Windows}\label{sec:processing_arrival}

We recognize that the main affective response in spatial navigation is during ``the experience of an arrival, the way in which you come into a space for the first time'' \cite{totten2014architectural}. We define \emph{arrival} as the transition from one space to the next, and form an \emph{arrival time window} around the time that a new room is entered in the video, including prior moments when the sliding door in the previous room opens and the new room is revealed. Therefore, considering $t_e$ as the timestamp (in seconds) that the player enters the new room's area, the arrival time window is between $[t_e-1,t_e+2]$ seconds (see Fig.~\ref{fig:room_window}). For this analysis, we are interested in the \emph{gradient} of the annotation trace during this 3-second arrival time window.
The gradient is calculated as the change between consecutive time frames \cite{yannakakis2018ordinal,yannakakis2017ordinal,camilleri2017generalmodels}. The gradient values of the entire trace are normalized to have an amplitude of 1 but the sign of the gradient is retained. Averaging the normalized gradient on the 3 seconds of the arrival window, we derive the \emph{arrival's mean gradient of arousal}.

\section{Results}\label{sec:results}

This section explores the impact of a change in spatial features on the annotators' arousal levels. Following \cite{yannakakis2018ordinal,yannakakis2017ordinal}, the analysis treats the annotators' traces and the design features in a \textit{relative} fashion, observing how changes in the design features of consecutive rooms impacts the mean arousal of the next room compared to the previous one (Section \ref{sec:results_changes}) and how the moment-to-moment arousal levels change during the arrival at a new room (Section \ref{sec:results_arrivals}).

\subsection{Changes in Mean Arousal}\label{sec:results_changes}

As a first experiment, we compare consecutive rooms in terms of change in each of the design features and in terms of change of the mean room arousal. Since room features are categorical, determining whether the feature changes is straightforward. In the 20 videos recorded (i.e. 23 transitions per video or 460 transitions in total), there are 219 changes in curvature, 238 changes in size, 325 changes in light color, and 243 changes in complexity between consecutive rooms. On the other hand, mean room arousal is a real value within $[0,1]$ and thus what constitutes a change between the mean arousal values of two consecutive rooms needs to be defined. Following the literature on treating affect as rankings \cite{yannakakis2017ordinal,yannakakis2018ordinal,makantasis2019pixels}, we determine an uncertainty threshold $\epsilon$ and if the absolute difference between mean arousal values of consecutive rooms is below this threshold then we consider that there is no change in arousal. For mean arousal, $\epsilon=0.05$, i.e. 5\% of the value range of each arousal trace. We consider that arousal increases from the previous room ($r_1$) to the next ($r_2$) if $m(r_2)-m(r_1)>\epsilon$, and that arousal decreases if $m(r_2)-m(r_1)<-\epsilon$. We then check those transitions from one room to the next where a specific design feature changes, and if there is an arousal increase or decrease then we mark it as an \textit{arousal shift}. If the design feature is absent in the previous room but is present in the next room (or the color warmth increases by at least one step) and arousal increases, then we mark it as \textit{agreement}; if arousal decreases we mark it as \textit{disagreement}. Similarly if the design feature is present in the previous room but is absent in the next room (or the color warmth decreases by at least one step) and arousal decreases, we mark \emph{agreement}, and if arousal increases we mark \emph{disagreement}. Through this simple process we enumerate the instances where the change in a design feature has a corresponding change in arousal. A high agreement ratio means that the presence of a design feature leads to higher arousal, while a ratio of arousal shifts over the total number of feature changes in transitions means that annotators have a reaction when this particular feature changes.

\begin{table}[!tb]
\centering
\caption{Changes in arousal annotations matching with room properties, per annotator and based on agreement between annotators. Significant agreements or disagreements are shown in bold. $N$ displays the number of datapoints (changes in mean arousal) remaining, also as ratio over all 460 transitions.}
\label{tab:results_mean_arousal}
\begin{tabular}{l|c|c|c|c}
 &  & Room & Color  &  \\
 & Curvature & Size & Warmth & Complexity \\ \hline \hline
\multicolumn{5}{c}{Annotator A ($N=413$, 90\%)}\\\hline
agreements & \textbf{58\%} & 55\% & 54\% & \textbf{97\%} \\ \hline
disagreements & 42\% & 45\% & 46\% & 3\% \\ \hline
arousal shifts & 91\% & 89\% & 89\% & 94\% \\ \hline
\hline
\multicolumn{5}{c}{Annotator B ($N=351$, 76\%)}\\\hline
agreements & \textbf{83\%} & \textbf{57\%} &  \textbf{56\%} & \textbf{60\%} \\ \hline
disagreements & 17\% & 43\% & 44\% & 40\% \\ \hline
arousal shifts & 81\% & 76\% & 76\% & 77\% \\ \hline
\hline
\multicolumn{5}{c}{Annotator C ($N=250$, 54\%)}\\\hline
agreements  & 49\% & 52\% & 50\% & 50\% \\ \hline
disagreements & 51\% & 48\% & 50\% & 50\% \\ \hline
arousal shifts & 57\% & 51\% & 53\% & 56\% \\ 
\hline \hline
\multicolumn{5}{c}{At least 2 annotators agree on room arousal change ($N=254$, {55\%})}\\\hline
agreements & \textbf{75\%} & 54\% & \textbf{59\%} & \textbf{87\%} \\ \hline
disagreements & 25\% & 46\% & 41\% & 13\% \\ \hline
 arousal shifts & 61\% & 56\% & 55\% & 57\% \\
\hline \hline
\multicolumn{5}{c}{All 3 annotators agree on room arousal change ($N=49$, 11\%)}\\\hline
agreements & \textbf{87\%} & 61\% & 51\% & \textbf{94\%} \\ \hline
disagreements & 13\% & 39\% & 49\% & 6\% \\ \hline
arousal shifts& 14\% & 10\% & 11\% & 15\% \\ 
\end{tabular}
\end{table}

Table \ref{tab:results_mean_arousal} shows the agreements between mean arousal changes and design feature changes, per annotator. It is evident that some annotators were less prone to shift their arousal annotations between rooms (e.g. observing the overall arousal shifts of Annotator C). On the other hand, Annotator B is fairly consistent and the presence of every spatial feature is more often associated with increased arousal than not. Significance of the ratio of agreements versus disagreements is calculated based on the binomial distribution of all arousal changes when the spatial feature changes, assuming a 50\% probability that the changes may be in agreement. Significance is established at 95\% confidence. It is evident that different annotators provide traces with different degrees of granularity, while some extremes (e.g. 97\% agreement in terms of complexity for Annotator A) raise some concerns discussed in Section \ref{sec:discussion}.

\subsection{Inter-rater Agreement on the Changes in Mean Arousal}\label{sec:results_interrater}

Based on their raw arousal traces, annotators are often in agreement (Crombach $\alpha=0.717$). However, we observe that patterns are less clear when we analyze how each annotator's trace seems to be impacted by changes in design features. We focus on those mean arousal changes where at least two annotators are in agreement (i.e. mean room arousal increases for at least two annotators, or decreases for two annotators). Calculating the instances where at least two annotators are in agreement, and matching them with changes between consecutive rooms, we retain 254 arousal shifts out of a total of 460 room transitions, which is a good sample for data analysis. Table \ref{tab:results_mean_arousal} includes the agreements, disagreements and arousal shifts with each spatial feature change in consecutive rooms for instances where at least two annotators agree. It is evident that higher complexity and higher curvature leads to higher arousal, with warmer colors also coinciding with high arousal to a significant degree. Moreover, changes in curvature are more likely to result in a non-trivial change in arousal. This is surprising, as we expected that features of color would have a more noticeable effect than all features of form. Looking into the impact of different color transitions on mean room arousal change in this corpus, we found significant influences when the player entered a blue room from any other color room (resulting in decreased arousal in 85\% of instances).

For completeness, we performed an analysis with those instances where all three annotators agreed in terms of mean room arousal changes. The results are included in Table \ref{tab:results_mean_arousal}. As expected, agreements are even more pronounced, however the number of arousal shifts where all three annotators have a non-trivial increase or decrease is 49 out of 460. This means that the findings are rather circumstantial, and we lose most of the data for the sake of inter-rater agreement.

\subsection{Impact of Arrivals}\label{sec:results_arrivals}

\begin{table}[!tb]
\centering
\caption{Agreement between sign of the arousal gradient and changes in room properties during an arrival. Significant agreements or disagreements are shown in bold. $N$ displays the  number of datapoints (non-zero arousal gradients) remaining, also as ratio over all 460 transitions.}
\label{tab:results_arrivals}
\begin{tabular}{l|c|c|c|c}
 &  & Room & Color  &  \\
 & Curvature & Size & Warmth & Complexity \\ \hline \hline
\multicolumn{5}{c}{Annotator A ($N=432$, 94\%)}\\\hline
agreements & \textbf{57\%} & 53\% & 53\% & \textbf{96\%} \\ \hline
disagreements & 43\% & 47\% & 47\% & 4\% \\ \hline
arousal shifts & 95\% & 95\% & 94\% & 98\% \\ \hline
\hline
\multicolumn{5}{c}{Annotator B ($N=347$, 75\%))}\\\hline
agreements & \textbf{84\%} & \textbf{58\%} & \textbf{56\%} & \textbf{60\%} \\ \hline
disagreements & 16\% & 42\% & 44\% & 40\% \\ \hline
arousal shifts & 80\% & 75\% & 75\% & 74\% \\ \hline
\hline
\multicolumn{5}{c}{Annotator C ($N=260$, 57\%)}\\\hline
agreements & 53\% & 50\% & 49\% & 47\% \\ \hline
disagreements & 47\% & 50\% & 51\% & 53\% \\ \hline
arousal shifts & 85\% & 90\% & 88\% & 90\% \\ \hline
\hline 
\multicolumn{5}{c}{At least 2 annotators agree on sign of arousal gradient ($N=309$, 67\%)}\\\hline
agreements & \textbf{73\%} & 51\% & \textbf{56\%} & \textbf{80\%} \\ \hline
disagreements & 27\% & 49\% & 44\% & 20\% \\ \hline
arousal shifts & 72\% & 69\% & 67\% & 69\% \\ \hline
\hline 
\multicolumn{5}{c}{All 3 annotators agree on sign of arousal gradient ($N=89$, 19\%)}\\\hline
agreements & \textbf{84\%} & 59\% & 59\% & \textbf{90\%} \\ \hline
disagreements & 16\% & 41\% & 41\% & 10\% \\ \hline
arousal shifts & 22\% & 21\% & 20\% & 20\% \\
\end{tabular}
\end{table}

As noted in Section \ref{sec:processing_arrival}, the point when the player enters a room (i.e. ``arrival'') is expected to have a strong impact on the affective response. Calculating the average gradient within the 3-second time window during this transition between rooms, we check when the mean arousal gradient is positive or negative (ignoring values between $10^{-4}$ and $-10^{-4}$ as ambiguous, determined empirically) matches the change in each design feature between the previous room (that the player exits) and the next room that they enter. Table \ref{tab:results_arrivals} shows each annotator's agreements between arrival gradients and changes in design features, as well as on the filtered data where gradients had the same sign for at least two or all three annotators. Unsurprisingly, results follow a similar pattern as with changes in mean arousal, with annotator B showing significant impact of all design features on arousal change during the arrival time window and annotator C being more ambiguous in their annotations. Notably, however, annotator C has clearer patterns than with mean arousal changes. Moreover, the number of instances where arousal gradient was non-zero when a feature changes is increased over the times mean arousal changes (see ``arousal shifts'' entries in Table \ref{tab:results_mean_arousal}). This may indicate that focusing on the arrival windows and their arousal gradient could provide more concise data, although we cannot discount other effects on data processing such as the different thresholding procedures for the two signals. Focusing on the instances where two annotators agree as the best consensus between sufficient data and inter-rater agreement, we observe that complexity and curvature of the space have a strong impact on arousal during the moments of arrival, with color warmth having a significant but less pronounced effect.

\section{Discussion}\label{sec:discussion}

This first study explored how both the transitions between spaces and arousal can be treated in a time-continuous fashion. Results of Section \ref{sec:results} indicate that the presence of curved forms or occlusion from complex structures leads to increased arousal. Moreover, by exploring the impact of blue, red, and neutral light we notice a more complex relationship between color and arousal. Treating warmth of the color as a linear variable, we see that warmer colors tend to lead to higher arousal; this is in line with extensive research on both cognitive psychology \cite{knez2001ffects,mccloughan1996lighting} and digital spaces \cite{niedenthal2008shadow,marples2020effect}. Finally, we have explored two different ways of processing the continuous annotation traces in a relative fashion, either looking at the mean arousal value per room and comparing consecutive rooms' mean arousal values, or focusing on a shorter time window of ``arrival'' and tracking the relative change of arousal within that time window. Both methods capture the change in arousal and juxtapose it with the changes in terms of each design feature to find matches and mismatches. As expected, both methods of processing the continuous annotation trace lead to similar conclusions, indicating that they are both valid for processing similar elicitors where the shift from one stimulus (in this case, room) to the next is gradual.

Along with experimental findings, the paper's contribution is the AffRooms dataset which consists of 62 minutes of footage of 3D spatial navigation and encompasses a variety of design features of architectural form and light color. The fact that each video includes all 24 possible combinations of features allows us to give annotators an inclusive view of the possible stimuli. The fact that the videos are pre-recorded ensures that navigational style is consistent and allows for the inter-rater agreement analysis in Section \ref{sec:results_interrater}. On the other hand, the long videos (each 3 to 4 minutes long) and the fact that annotators were not the ones in control of the navigation may have introduced some bias in the annotation traces. Due to the fact that (a) the spaces were often similar to each other (especially after the first few videos), (b) there was no goal or opposition to the player's navigation and (c) the annotators were watching another player navigate, it is possible that the annotations were based on cognitive rather than affective evaluation of the material. This seems to be true in the case of complex spaces for annotator A (see Table \ref{tab:results_mean_arousal}) where a change in arousal matched a change in complexity 97\% of the time: this indicates that early on the annotator developed a cognitive priming that complex spaces should be more arousing. The lack of purpose or context is also important: earlier studies \cite{marples2020effect} tasked players to navigate through mazes as quickly as possible, and thus the purpose of speed was explicit. In this case, the recorded navigation task aimed to provide a good view of each room while still keeping the duration manageable. In future work, more game-specific goals could be introduced when recording the videos, from e.g. finding a key inside the room in order to open the door to the next room, or avoiding patrolling enemies \cite{lopes2015horror}. 

The reported study was exploratory in nature, recruiting few annotators to perform thorough and homogeneous annotation which allowed us to calculate inter-rater agreements on the entire dataset. The annotation task was time-consuming, running slightly over an hour, and thus would likely not scale well for many annotators and modalities. A follow-up study will produce shorter navigation videos (with 12 room transitions instead of 24) and recruit many participants  via crowdsourcing platforms. Each participant would annotate a random subset of a video database. We also intend to explore other affect dimensions, such as valence, as time-continuous variables.

In terms of the elicitors, there is still a broad variety of features that can be explored; examples include sounds playing in each room, similar to \cite{lopes2015horror,marples2020effect}, walls' textures, more variants of complex rooms (e.g. with a central column or with clutter only along the edges of the room), etc. Among these, perhaps the easiest and most critical adaptation is the use of a more natural level of interior illumination, matching e.g. the Correlated Color temperature (CCT) of modern lightbulbs. The extreme saturation of blue and red light sources in the current experiment may explain why annotators were less consistent in terms of how light impacted arousal than for other features. While the colors used in this study are consistent with previous studies \cite{niedenthal2008shadow,marples2020effect}, it is worth exploring whether the difference between e.g. candlelight ($10^3$ Kelvin in CCT) versus daylight ($10^4$ Kelvin) has a clearer impact on arousal levels.

\section{Conclusion}\label{sec:conclusion}

Inspired by the many studies on the impact of architectural form and light on affect, this paper introduced the AffRooms corpus of 3D spatial navigation videos and explored how annotators reacted to changes across four design features. Unlike previous work, this experiment used an unbounded continuous annotation method which provided granular information on the moment-to-moment arousal shifts. Through this protocol, we were able to extract the relative changes in arousal from one room to the next, as well as the change in arousal at the moment of arrival at a new room. This initial study assumes only short-term memory on the part of the annotator and only compares the features of the previous room with the features of the next room. Results show that, although arousal traces are expectedly diverging across annotators, overall certain features of the 3D spaces strongly influence the arousal levels of the viewers. Future work should explore more complex ways of treating the signals (e.g. assuming a longer memory window and comparing all rooms against all rooms) and expand the dataset with more architectural features and real-life illumination.

\end{document}